# Selective metal deposition at graphene line defects by atomic layer deposition

Kwanpyo Kim[1,2,†], Han-Bo-Ram Lee[1,3,†], Richard W. Johnson[4], Jukka T. Tanskanen[1,5], Nan Liu[1], Myung-Gil Kim[1,6], Changhyun Pang[1,7], Chiyui Ahn[8], Stacey F. Bent[1], and Zhenan Bao[1,*]

[1]*Department of Chemical Engineering, Stanford University, Stanford, CA 94305, U.S.A.*

[2]*Department of Physics, Ulsan National Institute of Science and Technology (UNIST), Ulsan 689-798, Korea*

[3]*Department of Materials Science and Engineering, Incheon National University, Incheon 406-772, Korea*

[4]*Department of Materials Science and Engineering, Stanford University, Stanford, CA 94305, U.S.A.*

[5]*Department of Chemistry, University of Eastern Finland, Joensuu, Finland*

[6]*Department of Chemistry, Chung-Ang University, 221 Heukseok-dong, Dongjak-gu, Seoul 156-756, Korea*

[7]*School of Chemical Engineering, Sungkyunkwan University (SKKU), Suwon 440-746, Korea*

[8]*Department of Electrical Engineering, Stanford University, Stanford, CA 94305, U.S.A.*

* To whom correspondence should be addressed: zbao@stanford.edu

† These authors contributed equally to this work.

**One-dimensional defects in graphene have strong influence on its physical properties, such as electrical charge transport and mechanical strength. With enhanced chemical reactivity, such defects may also allow us to selectively functionalize the material and systematically tune the properties of graphene. Here we demonstrate the selective deposition of metal at chemical vapour deposited graphene's line defects, notably grain boundaries, by atomic layer deposition. Atomic**

**layer deposition allows us to deposit Pt predominantly on graphene's grain boundaries, folds, and cracks due to the enhanced chemical reactivity of these line defects, which is directly confirmed by transmission electron microscopy imaging. The selective functionalization of graphene defect sites, together with the nanowire morphology of deposited Pt, yields a superior platform for sensing applications. Using Pt-graphene hybrid structures, we demonstrate high-performance hydrogen gas sensors at room temperatures and show its advantages over other evaporative Pt deposition methods, in which Pt decorates graphene surface non-selectively.**

**Introduction**

Graphene, an atomically thick $sp^2$-bonded carbon membrane, has excellent electrical and optical properties such as high charge carrier mobility[1] and tunable optical absorption[2]. Owing to these excellent properties, graphene has emerged as a promising candidate in applications such as optoelectronics[3] and high-frequency electronics[4]. Graphene also has promising properties towards high speed and ultra-sensitive gas/vapor sensors because every atom in graphene is a surface atom and it can readily respond to environmental changes[5-7]. For successful applications, large-area high-quality graphene growth has been intensively studied, and significant advances have recently emerged[8-11]. Among them, chemical vapor deposition (CVD) has achieved successful growth of high-quality, polycrystalline monolayer graphene on metals[9,12]. Although it can be generally of good quality, CVD graphene still has numerous synthesis-related defect structures, especially one-dimensional (1D) line defects. Typically, CVD graphene synthesis yields a polycrystalline graphene structure with grain sizes in the order of a few micrometers[9,13-15]. Moreover, the transfer process of graphene inevitably renders wrinkles[16-18], folds[19], and cracks[20] in CVD graphene. These line defects in CVD graphene can play important roles in determining CVD graphene's electrical and mechanical properties, as shown in recent studies[14,15,21-23].

The chemical properties of graphene defects are also of great interest. Generally, the pristine graphene lattice site is rather chemically inert; however, the introduction of atomic-scale defects in graphene can significantly modify its chemical and magnetic properties[24,25]. Previous theoretical studies shed light on enhanced chemical reactivity of 1D line defects, especially grain boundaries, in CVD graphene[26-28] but the available experimental study on this subject is limited. The chemical functionalization at these chemically reactive sites would be an interesting way to modify the physical and chemical properties of CVD graphene.

In this paper, we demonstrate the selective deposition of metal at the line defects of polycrystalline CVD graphene, notably grain boundaries, via atomic layer deposition (ALD). ALD allows us to deposit Pt predominantly at graphene's grain boundaries, folds, and cracks due to the enhanced chemical reactivity at these defect sites. Transmission electron microscopy (TEM) directly confirms that most of the Pt deposition sites coincide with graphene line defects. By density functional theory (DFT) calculations, we confirm that a graphene grain boundary has higher chemical reactivity compared to pristine lattice due to local strained C-C bonding. As the number of ALD cycles increases, Pt nanoparticles deposited along the defect sites coalesce and adopt nanowire morphology. The optical transmittance and electrical conductivity of the obtained graphene-Pt hybrid structures are studied as a function of Pt deposition. Finally, we demonstrate high-performance hydrogen gas sensors using Pt ALD graphene hybrid structures and show its advantages over evaporative Pt deposition methods, in which Pt decorates the surface of graphene nonselectively, regardless of defect locations. The selective functionalization of graphene defect sites, with the nanowire morphology of deposited Pt, yields a superior platform for sensing applications.

**Results**

**Selective deposition of Pt at graphene line defects and TEM characterizations.** Previous ALD studies on carbon nanotubes and graphene have focused on the deposition of high-κ dielectric materials for transistor applications. For this purpose, uniform deposition of the dielectric material is desirable. Without surface functionalization, the carbon nanotube and graphene surface is inert and the metal oxide will grow only at defect structures such as graphite step edges (or graphene edges)[29-31]. On the other hand, surface functionalization can be used to deposit a uniform metal oxide film on carbon nanotube and graphene surface[29,32-34]. Metal deposition by ALD has recently gained much attention[35-37] and, notably, selective metal (Pt) growth at the step edges of highly ordered pyrolytic graphite (HOPG) also has been demonstrated[38]. The HOPG step edges serve as Pt nucleation sites due to their higher chemical reactivity compared to the inert basal plane, which arise from a combination of dangling bonds and functional groups. A similar concept can be applied to CVD graphene. When the ALD process is utilized, metal will be selectively deposited at the 1D line defects in CVD graphene due to their enhanced chemical reactivity. Figure 1a shows a schematic of selective metal growth at 1D defect sites of polycrystalline CVD graphene by ALD. From this process, we expect to obtain a graphene-metal hybrid structure, where a network of metal nanowires decorates graphene line defects. Such a structure may be of interest for transparent electrodes and sensors for which the control of conductivity across grain boundaries is important.

We synthesize polycrystalline CVD graphene using previously-reported growth conditions and transfer graphene to a glass slide with a poly (methyl methacrylate) (PMMA) support[9]. On these prepared graphene samples, we perform Pt ALD and systematically study the evolution of Pt deposition as a function of the number of ALD cycles. (See Method Section for the detailed sample preparation and Pt ALD process.) After the Pt deposition, SEM imaging is utilized to check Pt deposition on CVD graphene (Figure 1b,c). Figure 1b shows CVD graphene after 500 cycles of Pt ALD. The Pt-deposited sites are predominantly arranged in a nanowire shape. The distance between these line features are on the order of micrometers, which is on the same order of the graphene grain size in CVD graphene[13,19]. Additional ALD cycles give higher Pt coverage on CVD graphene. Figure 1c shows the Pt deposition on graphene after 1000 ALD cycles. We note that, other than line features, there are isolated Pt particles in CVD graphene samples. The isolated Pt particle deposition can be mainly attributed to Pt nucleation at point defects in prepared graphene samples. Possible surface contamination (mainly PMMA residues) can be eliminated from the main source of point Pt nucleation since controlled experiments with PMMA covered graphene do not induce Pt nucleation on surface with an identical ALD process (See Supplementary Figure 1).

Even though SEM images confirm the line-shaped morphology of ALD Pt on CVD graphene, it does not give us spatially resolved information on whether the Pt deposition sites correspond to the graphene defects. To directly confirm the selective growth of Pt at the graphene line defects, we employ transmission electron microscopy (TEM). After preparing suspended graphene TEM samples, we perform Pt ALD for TEM characterization with similar ALD conditions. Figure 2 shows the suspended graphene sample on a holey carbon TEM grid. The sample area outside the circles has a carbon support while inside the circle is the freestanding graphene (Fig. 2a).

The TEM image in Figure 2a confirms the distinct growth of Pt along line features with scattered point Pt nucleation, which is consistent with SEM imaging. It also confirms that the line features along which Pt growth occurs correspond to grain boundaries in the graphene. We acquire images around Pt line decoration at higher magnifications to obtain graphene lattice information as shown in Fig. 2b-d. The fast Fourier transforms (FFTs) of these images reveal that graphene has a relatively rotated lattice across the line features, confirming the presence of graphene tilt grain boundaries (Fig. 2e-g). Additional TEM images obtained in other areas also confirm that many of the Pt line features are originated from grain boundaries in CVD graphene. (See Supplementary Figures 2-4) We can also identify the rotational mismatch angles of each graphene grain from FFT analysis as shown in Figure 2a.

We compare Pt nuclear densities at the graphene grain boundaries and inside the grains. Even though we observe preferential growth of Pt at line defects, we also observe scattered Pt nucleation inside the grains, possibly due to graphene point defects or residues on the graphene surface. With TEM images of 300 ALD cycle samples, we find that the average distance between Pt nucleation is ~ 11 nm at graphene grain boundaries. If we assume the effective width of grain boundaries as 1 nm[13,14], this value corresponds to the Pt nucleation density of $8.8 \times 10^{-2}$ nm$^{-2}$. On the other hand, we find that the Pt nucleation density is $2.0 \times 10^{-4}$ nm$^{-2}$ inside the graphene grains. This is 440 times lower value compared to that at the grain boundaries, which clearly demonstrates the preferential Pt nucleation effect at the graphene grain boundaries. Raman spectroscopy can be used to estimate the defect density in graphene sample[39] and the observed Pt nucleation density inside the grains, $2.0 \times 10^{-4}$ nm$^{-2}$, is consistent with an estimated defect density in a high quality graphene sample (Supplementary Figure 5 and Supplementary Note 1).

Atomic resolution TEM imaging can be also performed at these functionalized graphene grain boundaries. Figure 3a clearly shows Pt decoration along a line feature after 300 ALD cycles. Magnified images at the upper and lower parts of the graphene lattice show that the two areas have a misaligned graphene lattice, also confirming the presence of a tilt grain boundary at the Pt line decoration. In that area, the upper (Fig. 3b) and lower (Fig. 3c) parts of the graphene lattice have a relative misorientation of 13 degrees. Pt nanoparticles have a diameter around 10 nm and some particles are successfully inter-connected. No visible hole is observed in graphene around grown Pt nanoparticles and suspended graphene sample maintains its structural integrity even after 1000 ALD cycles. Together with Raman spectroscopy measurements, these observations support that the utilized ALD process does not introduce extra defects in graphene. We also confirm that the selective Pt growth occurs on other kinds of graphene line defects, such as graphene folding structure and cracks (Supplementary Figures 6 and 7).

In terms of chemical reactivity, graphene grain boundaries are believed to have lower reactivity compared to step edges in graphite. For a step edge in HOPG, the chemically active dangling bonds are easily oxidized under exposure to the $O_2$ counter reactant, leading to nucleation of Pt on those sites[38]. On the other hand, an ideal graphene grain boundary composes of arrays of pentagon-heptagon carbon rings, where no obvious dangling bond is available[13,14,40-42]. After 300 ALD cycles, we find that the average size of Pt particles at grain boundaries is around 10 nm (Supplementary Figure 8), which is significantly smaller than the expected size based on the previous reported growth rate (39 nm with the growth rate of 1.3 Å/cycle)[38]. This can be attributed to the nucleation delay effect[38] and we estimate that the nucleation delay is around 200 cycles at the grain boundaries from the observed average particle size. Previously, Pt ALD on the step edges of HOPG have shown a nucleation delay of about 100 cycles[38]. The observed longer nucleation delay at the grain boundaries is an indirect evidence that a grain boundary has somewhat lower chemical reactivity compared to the step edges of HOPG.

**Density functional theory calculations of chemical reactivity at graphene grain boundaries.** To study the chemical reactivity of grain boundaries in detail, we perform density functional theory (DFT) calculations utilizing the PBE/PAW functional[43,44] (see Methods for details). The enhanced reactivity of grain boundaries has been theoretically explored in a number of previous studies[26-28] but energetic calculations for binding ALD precursors on graphene grain boundaries are not reported. Detailed energetic calculations can provide a point of comparison on binding different precursors, including various metal and oxide precursors, on graphene grain boundaries. We simulate a graphene grain boundary using a previously reported periodic grain boundary model with a single pentagon-heptagon pair[42], which is also confirmed by recent atomic-scale experimental observations[13,14]. We compare DFT-calculated energetics of the surface species formed as a product of the reaction $C_2^*$ + $Pt(CH_3)_3CpCH_3$ → $C-Pt(CH_3)_2CpCH_3^*$ + $C-CH_3^*$, where '*' refers to a surface species and 'Cp' to a cyclopentadienyl ring. Notably, it has been previously proposed in the context of Pt ALD on $TiO_2$, $Al_2O_3$, and $SrTiO_3$ that $Pt(CH_3)_3CpCH_3$ reacts on the surfaces by forming $Pt(CH_3)_2CpCH_3$ surface groups[45].

Three different $C_2^*$ reaction sites located at a strained region, i.e. at the pentagon-heptagon pair, are included in the study (see Figure 4a) and the results are illustrated in Figure 4. The PBE/PAW-calculated energetics suggest the Pt precursor chemisorption to a $C_2^*$ unit labelled by B and C ('B-C' site) in Figure 4 to be clearly favoured over the other sites. The high strain associated with the C-C (atoms labelled B and C in Figure 4) bond shared by the hexagon–heptagon unit is responsible for higher reactivity, in agreement with previous studies[26,28,46]. The reaction energy originates from the breaking of the Pt-$CH_3$ bond, the formation of C-Pt and C-$CH_3$ bonds on the surface, and the release of strain in the grain boundary[26,28] due to buckling of the carbon framework, as observed in all calculations with the Pt precursor bound at the pentagon-heptagon pair. On the other hand, the reaction product was found unstable on graphene basal plane (see Methods for details), demonstrating the higher reactivity of graphene grain boundary as compared to basal plane.

We also perform reaction energy calculations using a cluster model with a hybrid functional, PBE0 for various ALD precursors including $Pt(CH_3)_3CpCH_3$ and trimethylaluminum (TMA). (See

Methods and Supplementary Figures 9 and 10.) With $Pt(CH_3)_3CpCH_3$ reaction energy calculations, we find that the energetic trends from the cluster calculations are in agreement with the periodic calculations while the absolute values of the reaction energies are lower with the cluster model calculations (Supplementary Figure 9). We attribute this to the use of different functionals since hybrid functionals, such as PBE0, typically provide more accurate reaction energies than the GGA functionals. The different local strain fields in the two structure models can be also partially responsible for the different calculated reaction energies. The calculated somewhat unfavorable (positive) reaction energies for binding the Pt precursor on graphene grain boundaries are also consistent with the observed longer nucleation delay of around 200 cycles. Moreover, the TMA calculations suggest binding TMA on graphene grain boundaries is more difficult with respect to $Pt(CH_3)_3CpCH_3$, and hence TMA may be deposited preferably on more reactive sites on graphene, such as on graphene cracks and edges, compared to grain boundaries (Supplementary Figure 10). We note that a grain boundary reaction with $O_2$ reactant, $C_2^* + O_2 \rightarrow \text{C-O}^* + \text{C-O}^*$, can be also relevant to the ALD reaction process and enhanced reactivity with oxygen adatoms at graphene boundaries has been recently reported[28].

**Optical transmittance and sheet resistance measurements.** Having demonstrated the growth of Pt on graphene grain boundaries by ALD, we now turn to the influence of the deposited Pt on the electronic and optical characteristics of graphene. The monolayer CVD graphene sample has a sheet resistance of $R_s \geq \sim 1$ k$\Omega$/□, which is still a high resistance for successful application as a conducting transparent electrode[47]. Some origins of the degraded performance of CVD graphene conductance compared to values reported from high-quality monolayer graphene[48] comes from the one-dimensional defective structures of graphene, which can impede charge transport in graphene and significantly increase the sheet resistance. With respect to this, recent studies have demonstrated that, using a graphene-metal nanowires hybrid structure, very low sheet resistance can be achieved[49-51]. Similarly, Pt deposition on graphene can boost the electrical conduction since the Pt metal allows for an extra conduction channel. Moreover, since Pt ALD can leave most of the graphene area free of Pt deposition, the obtained graphene-Pt hybrid structure will be able to maintain high transmittance at optical light frequency.

We measure the optical transmittance and conductivity of graphene-Pt hybrid structures as a function of Pt deposition. The pristine graphene samples shows the flat optical transmittance with $T = 97.7$ % at 550 nm wavelength, which is consistent with the result from monolayer graphene sample [12] (Figure 5a.) Figure 5a also shows that the graphene samples with 500 and 1000 ALD cycles exhibit transmittance around 90 % and 60 %, respectively. We find that the Pt deposition on CVD graphene has some variation; the samples with 500 and 1000 ALD cycles have $T = 91 \pm 4$ % and $61 \pm 11$ %, respectively. We attribute this to the batch-to-batch variation in graphene defect density and the degree of surface cleanliness. We also find that the drop in optical transmittance is highly nonlinear to the number of ALD cycles; the formation of Pt islands, as we observed in SEM and TEM images, results in a nonlinear increase in Pt deposition area with respect to the number of ALD cycles.

The conductivity of graphene-Pt structure is also measured. We obtain the sheet resistance of pristine graphene samples transferred to glass slide around 1 k$\Omega$/□ in ambient conditions. As the Pt deposition proceeds, the graphene sheet resistance starts to decrease and reaches ~ 230 $\Omega$/□ at $T = 75$ % with 1000 ALD cycles (Fig. 5b). For samples with 500 ALD cycles ($T \sim 90$ %), the conductivity improvement is not significant compared to graphene samples without Pt deposition. Even though Pt particles are deposited at the grain boundaries adapting Pt nanowire morphology with 500 ALD cycles, poor inter-particle connectivity and particle-graphene interface may limit the expected conductivity improvement by Pt deposition (Fig. 1 and 2).

**H$_2$ sensing experiments with Pt ALD samples.** Graphene and carbon nanotubes, especially coupled with other metal and semiconducting materials, have been invested for gas sensing applications[6,7,52-55]. With Pt ALD graphene hybrid structures, we perform hydrogen gas sensing experiments. Functionalization of line defects can allow enhanced response for sensing applications[55] because this method selectively manipulates various linear defects of CVD graphene, which are the most sensitive sites related to its charge transport properties[14,15]. In this respect, Pt ALD on graphene can yield a superior platform for sensing applications, allowing the selective functionalization of graphene defect sites together with the nanowire morphology of deposited Pt.

To assess the effect of selective functionalization of graphene defects, we compare Pt-deposited graphene samples prepared by ALD and e-beam evaporation processes. E-beam evaporation is a physical deposition process and metal film forms via physisorption on graphene. With the limited mobility of Pt atoms during deposition, Pt islands are quite uniformly located even on the basal plane, regardless of defect locations. (Supplementary Figures 11 and 12).

We perform gas sensing experiments with different Pt thicknesses, since the gas sensing performance can be influenced by the coverage of Pt. Figure 6a shows the normalized resistance changes of various Pt-graphene samples responding to 0.5 % concentrated H$_2$ in N$_2$ gas. With the exposure to hydrogen gas (20 minutes), we observe a resistance increase for all the Pt-deposited graphene samples. The observed resistance increase is consistent with previous reports[53,56-59]. The mechanism of hydrogen gas sensing is mainly ascribed to the change of doping levels of graphene. Previous investigations have found that the hydrogen molecules dissociate into atomic hydrogen on a Pt surface and the resulting atomic hydrogen lowers the work function of Pt. This in turns cause the electron transfer from Pt to graphene and reduces the p-doping levels of graphene, resulting in the increase of the resistance[53,56-59].

The sample prepared by e-beam evaporation shows an optimal thickness for gas sensing response[60]. This is related to the coverage of metals on graphene surface. Ideally, higher Pt coverage on graphene allows for a greater amount of reaction. However, as thicker Pt films are deposited, the Pt metal will eventually form a separate metallic conduction path, which is not sensitive to hydrogen exposure and therefore reduces the sensing performance. In our case, graphene samples on a regular glass slide show an optimal Pt thickness of 0.5 nm for the gas sensing application (Fig 6a). 0.5 nm Pt deposited on graphene shows a normalized response change of around 50 % after 20 minutes of exposure to 0.5 % hydrogen gas. With a higher deposition of Pt, 3 nm, the sheet resistance of graphene samples are significantly reduced from ~ 1 kΩ/□ to ~ 240 Ω/□ (Supplementary Table 2) and metallic conduction path from Pt film is dominant. Therefore, the normalized resistance change to hydrogen exposure shows a significantly reduced response.

Compared to the best-performing samples by e-beam evaporation (0.5 nm Pt on graphene), the Pt-graphene sample prepared by ALD (1000 cycles) clearly shows a better gas sensing performance. We observe that the 1000 Pt ALD cycle sample shows a normalized resistance change around 100 % with 0.5 % hydrogen gas exposure. Moreover, the response to hydrogen gas shows a much faster response compared to evaporated samples, as shown in Fig. 6a. We study the performance of 1000 Pt ALD cycle samples in further detail. We measure the sensing response with the 1000 Pt ALD cycles sample to an ultra-low concentration of $H_2$ gas of 2 ppm. As shown in Figure 6b, we clearly observe a resistance change upon exposure to 2 ppm hydrogen gas. Figure 6c shows the normalized resistance changes of the same sample responding to various hydrogen concentrations. The normalized resistance change after 20 minutes of hydrogen exposure shows a clear concentration dependence as shown in Figure 6d. We also plot the initial rate of normalized resistance change as another sensing parameter[61], because the resistance of the samples shows the non-saturating response even after 20 minutes of exposure for low concentration. Both of the resistance change and the rate of resistance change display very similar concentration-dependent responses. The responses also start to deviate from a straight line, showing saturation above around 500 ppm of hydrogen concentrations.

**Discussion**

We attribute the observed enhanced gas sensing performance of ALD samples to two main mechanisms. First, the Pt decoration adapting nanowire shape at the grain boundaries by ALD can provide better functionalization geometry for enhanced electrical response, compared to isolated particle decorations by e-beam evaporation. With Pt nanowires at the grain boundaries, a conduction path between electrodes inevitably goes through Pt-functionalized regions, which are the area of resistance increase via local graphene doping changes[53,59] or carrier depletion[55] upon gas exposure. On the other hand, with isolated particle decorations, one can find a conduction pathway, which goes through non-affected regions (Supplementary Figure 13). If we assume an extreme case where the local resistance increase is infinite by carrier depletion[55], the conduction pathway will be totally blocked with linear functionalization while the conduction is still possible with point decoration; the effect of resistance change is bigger for linear functionalization. A simple resistance model also predicts the overall sample resistance increase which scales linearly with a local resistance increase for the linear functionalization case whereas the overall resistance rises sub-linearly for the point functionalization. Therefore, the resistance increase will have enhanced effect in the case of linear decoration with comparable Pt decoration coverage (Supplementary Figure 13). As a second mechanism for the enhanced sensing response, the interaction between graphene defect sites and dissociated atomic hydrogen may also play important role, as discussed in carbon nanotube defect functionalization and related sensing measurements[55].

In conclusion, we demonstrate selective Pt growth at graphene defect sites via ALD. Through direct TEM investigation, we clearly show that Pt predominantly grows at graphene's line defect sites, such as grain boundaries. Since ALD allows for the selective growth of materials on defects of CVD graphene, this method can be used to visualize the locations of graphene defects[62] and obtain important information on graphene samples, such as grain size of the graphene sample[63]. We demonstrate that the metal-graphene hybrid structure obtained by ALD can perform as a high-performance hydrogen gas sensor owing to the unique selective functionalization of graphene line defects. Moreover, the metal deposition by ALD is not limited to Pt. The deposition of metals with higher electrical conductivity, such as Ag and Au, would also aid obtaining lower sheet resistance for transparent conducting electrode applications and give other functionality for sensing and energy storage applications[7].

## Methods

**Graphene sample preparation.** Graphene was synthesized by chemical vapour deposition (CVD) on 25 μm thick copper foil (99.8% Alfa Aesar, Ward Hill, MA)[9]. In brief, copper foil was inserted into a quartz tube and heated to 1040 °C with flowing 10 sccm $H_2$ at 100 mTorr. After annealing for 1 hour, the gas mixture of 25 sccm $CH_4$ and 5 sccm $H_2$ at 450 mTorr was introduced for 20 min to synthesize graphene. Finally, fast cool to room temperature with flowing 25 sccm $CH_4$ and 5 sccm H2 was performed.

**Graphene transfer to glass slide.** For SEM, Raman spectroscopy, optical transmittance, and sheet resistance measurement characterization, graphene was transferred to glass slides with PMMA support[9]. A PMMA solution (poly (methyl methacrylate), average Mw ~996 000 by GPC, Sigma-Aldrich product no.182265, dissolved in chlorobenzene with a concentration of 46 mg/ml) was spin-coated on the surface of as-grown graphene on Cu foil at the speed of 2000 rpm for 1 min. The sample was left in air for one day to allow the solvent evaporate thoroughly. $O_2$ plasma was then applied to remove the graphene layer on the other side of the Cu foil. Then the sample was placed into a solution of sodium persulfate ($Na_2S_2O_8$, a concentration of 0.1 g in 1 mL of water) to etch the underlying copper foil and is then rinsed with deionized water. The PMMA/graphene films were picked up by glass slides and left for 24 hours to obtain completed dry samples. The PMMA film was removed by soaking in aceton for 24 hours and then rinsed with isopropyl alcohol followed by blow dry. Finally, annealing samples with $H_2$ (20 %) and Ar (80 %) environment (with total pressure ~ 1 Torr) at 360 °C for 2 hours was performed to remove residual PMMA and obtain cleaner graphene surface.

**Graphene TEM grid preparation.** For TEM characterization, graphene was transferred to Quantifoil holey carbon TEM grids (SPI Supplies, 300 meshes, 2 μm hole size) using a direct transfer method[13,64]. We placed the TEM grid onto a graphene-covered copper foil with carbon film side facing the graphene. Then a small amount of isopropyl alcohol (IPA) was dropped on to the sample and air-dried. Additional flattening (prior to IPA step) of copper foil or TEM grid by sandwiching between glass slides ensures better adhesion. Finally, the sample was placed into a solution of sodium persulfate ($Na_2S_2O_8$, a concentration of 0.1 g in 1 mL of water) to etch the underlying copper foil and is then rinsed with deionized water.

**Atomic layer deposition of Pt.** Pt was deposited in a custom designed warm-wall ALD reactor controlled by LabVIEW software by using a metalorganic Pt precursor (methylcyclopentadienyltrimethyl-platinum) and air counter-reactant. The graphene surface was pulsed with precursor and counter-reactant for 2 s (50 mTorr and 500 mTorr, respectively) during each step and the chamber was purged for 8 s between pulses. The substrate temperature was fixed at 300 °C. More information on the chamber apparatus and deposition conditions can be found elsewhere[38].

**Aberration-corrected TEM imaging.** TEM imaging was performed at 80 kV using a FEI Titan equipped with a spherical aberration (Cs) corrector in the image-forming (objective) lens and a monochromator. The $C_s$ coefficient was set to approximately -10 μm. The images were acquired using an Ultrascan 1000 CCD camera.

**Characterization.** Scanning electron microscopy (SEM) images were acquired with a FEI XL30 Sirion SEM with a FEG source, operated at 5 kV. Extra TEM imaging and EDX measurement were performed with a FEI Tecnai F20 operated at 200 kV. Raman spectra were measured using a WiTech confocal Raman microscope which is equipped with a piezo scanner and an intensity-tunable 532 nm NiYAG laser. The optical transmittance measurement was performed with an Agilent Cary 6000i UV/Vis/NIR spectrometer.

**Conductivity measurement and hydrogen gas sensing experiment.** Graphene sheet resistance measurement was performed by a four-probe measurement (parallel Au electrodes on graphene) with a Keithley 4200-SCS. $H_2$ gas sensing experiment was performed with a home-built measurement set-up.

The $H_2$ concentration was controlled by mass flow controllers and the four-probe resistance of samples was recorded (Supplementary Figure 14).

**Computational details.** The DFT calculations were performed by using the Vienna Ab initio Simulation Program (VASP)[43,44] and the projector-augmented-wave (PAW) method. A periodic grain boundary model[42], corresponding to a (1,0) dislocation, with a unit cell composition of $C_{292}$ was utilized. The periodic structural relaxations were performed using the GGA functional PBE[65] with the PAW potentials implemented in VASP, the conjugate gradient algorithm (IBRION = 2 in VASP input), and Γ point k-sampling. Default accuracy parameters for fast Fourier transform (FFT) grid and real space projectors (PREC = NORMAL in VASP input) were adopted. A vacuum thickness of 20 Å was utilized to avoid interactions between graphene and its periodic images.

Supplementary calculations were performed by using the PBE0[65,66] functional with the standard split-valence + polarization (def-SVP)[67,68] basis set and without symmetry constraints. Quasirelativistic effective core potentials (ECP) were utilized for 60 core electrons of Pt[69]. A $C_{50}H_{16}$ cluster with H-terminated edges and a diameter of about 1.5 nm was utilized to represent the local atomic structure of graphene grain boundary with pentagon-heptagon pairs. The graphene basal plane was simulated by a $C_{42}H_{16}$ cluster with H-terminated edges. Full structural optimizations without symmetry constrained by the PBE0 hybrid functional were performed on the pristine graphene grain boundary cluster, the $Pt(CH_3)_3CpCH_3$ molecule, and the systems with $Pt(CH_3)_3CpCH_3$ chemisorbed on the grain boundary cluster. The computed total energies were used in the determination of the reaction energies (Supplementary Figure 9). In the investigation of $Pt(CH_3)_3CpCH_3$ chemisorption on the $C_2^*$ reaction sites on the graphene grain boundary, both structural alternatives of the product were considered and the energetically favoured one was reported. All cluster calculations were performed with Gaussian09 software package[70].

For comparison with the results for binding $Pt(CH_3)_3CpCH_3$ on graphene grain boundaries, we performed analogous cluster calculations for trimethylaluminum (TMA), which is a typical ALD precursor for depositing alumina. The results are summarized in Figure S10, where the sites B-C, C-C, and C-D are energetically practically equal for TMA chemisorption. The stabilization of these sites with respect to the A-B site is due to coordination of the Al atom to two neighboring C atoms with the Al atom being located above the C-C bond. In comparison with the $Pt(CH_3)_3CpCH_3$ calculations, the most favorable reaction energies for TMA are around 15 kcal/mol, whereas the calculated reaction energy for binding $Pt(CH_3)_3CpCH_3$ on the B-C site is about 8 kcal/mol. Thus, the results suggest binding TMA on graphene grain boundaries is more difficult with respect to $Pt(CH_3)_3CpCH_3$ and, hence, TMA might only bind on more reactive sites on graphene, such as on graphene cracks and graphene edges.

**Figures**

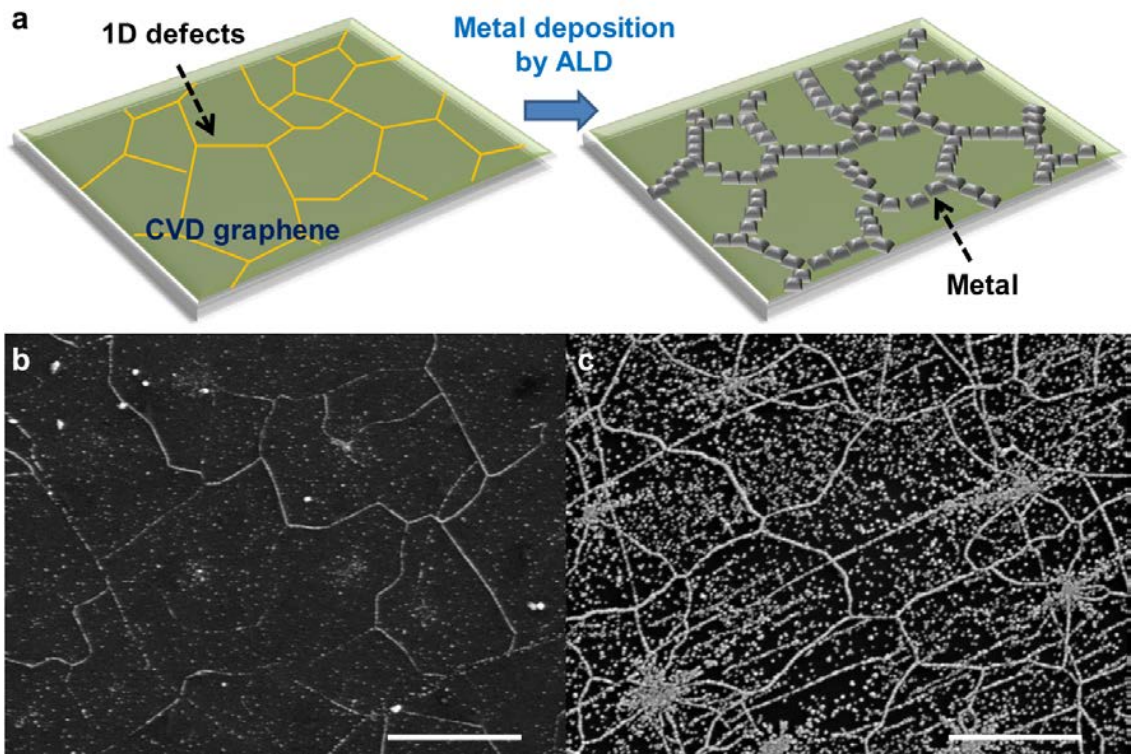

**Figure 1. Selective Pt growth by ALD on one-dimensional (1D) defect sites of polycrystalline CVD graphene.** (**a**) Schematic of selective Pt growth on 1D defects in CVD graphene. Various line defects, such as grain boundaries, cracks, and folded structures are present in CVD graphene. By utilizing atomic layer deposition, metal can be selectively deposited at the one-dimensional defect sites in graphene. The graphene-metal hybrid structure can be obtained through this process. (**b**) SEM images of CVD graphene on a glass substrate after 500 ALD cycles of Pt deposition. Pt growth shows the predominant line shape. Scale bar, 2 µm. (**c**) SEM images of CVD graphene with Pt deposition after 1000 ALD cycles. Scale bar, 2 µm.

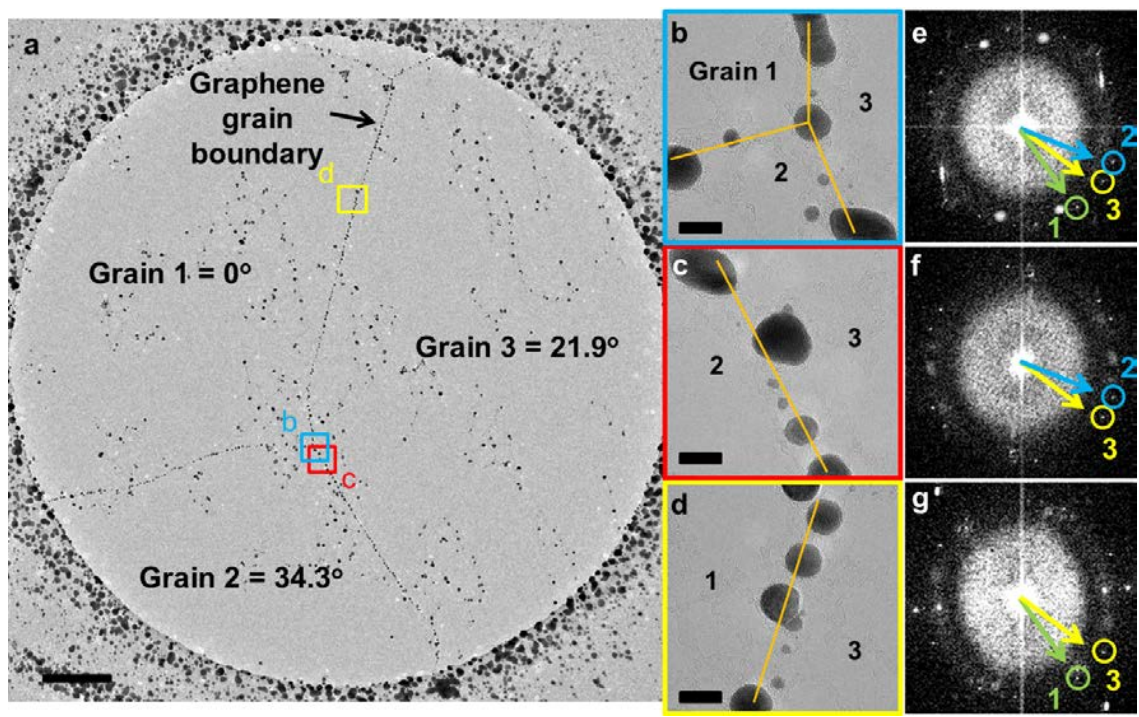

**Figure 2. TEM images of Pt growth along grain boundary of CVD graphene.** **(a)** TEM image of suspended CVD graphene after 300 cycles of Pt ALD growth. The distinct growth of Pt along line features is observed. Graphene is supported by an amorphous carbon film surrounding the area of the circle; the enhanced Pt growth seen in this region is occurring on the amorphous carbon TEM support. The small coloured boxes are the field of views for figure b, c, and d. The relative misorientation angle (with respect to grain 1) of graphene grains is shown for each grain. Scale bar, 200 nm. **(b)** Magnified TEM image at the triple junction of graphene grain boundary. The yellow lines show the location of graphene grain boundaries where the predominant Pt growth is observed. Scale bar, 10 nm. **(c-d)** Magnified TEM images around grain boundaries at different areas. Scale bar, 10 nm. **(e)** The Fourier transform of image b. The circled diffraction signals originate from three misoriented graphene grains. **(f)** The Fourier transform of image c. It shows two sets of graphene diffraction patterns from grain 2 and 3. **(g)** The Fourier transform of image d, showing two sets of graphene diffraction patterns from grain 1 and 3.

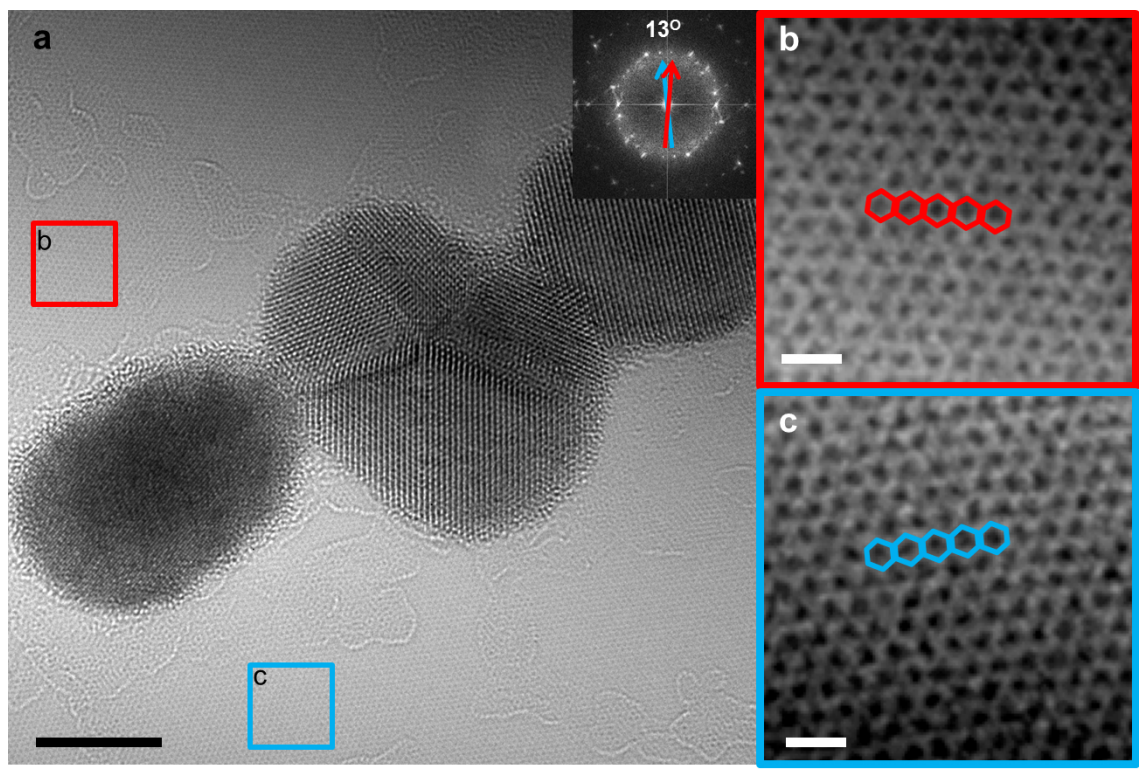

**Figure 3. Atomic resolution TEM image of Pt growth at graphene grain boundary.** **(a)** Atomic resolution TEM image at graphene grain boundary with Pt decoration after 300 ALD cycles. The small coloured boxes are the field of view for figure b and c. The inset shows the Fourier transform of the image. Scale bar, 5 nm. **(b)** The zoomed-in image at the box b. Scale bar, 5 Å. **(c)** The zoomed-in image at the box c. The upper (b) and lower (c) parts of the graphene lattice have a relative rotation of 13 degrees. Scale bar, 0.5 nm.

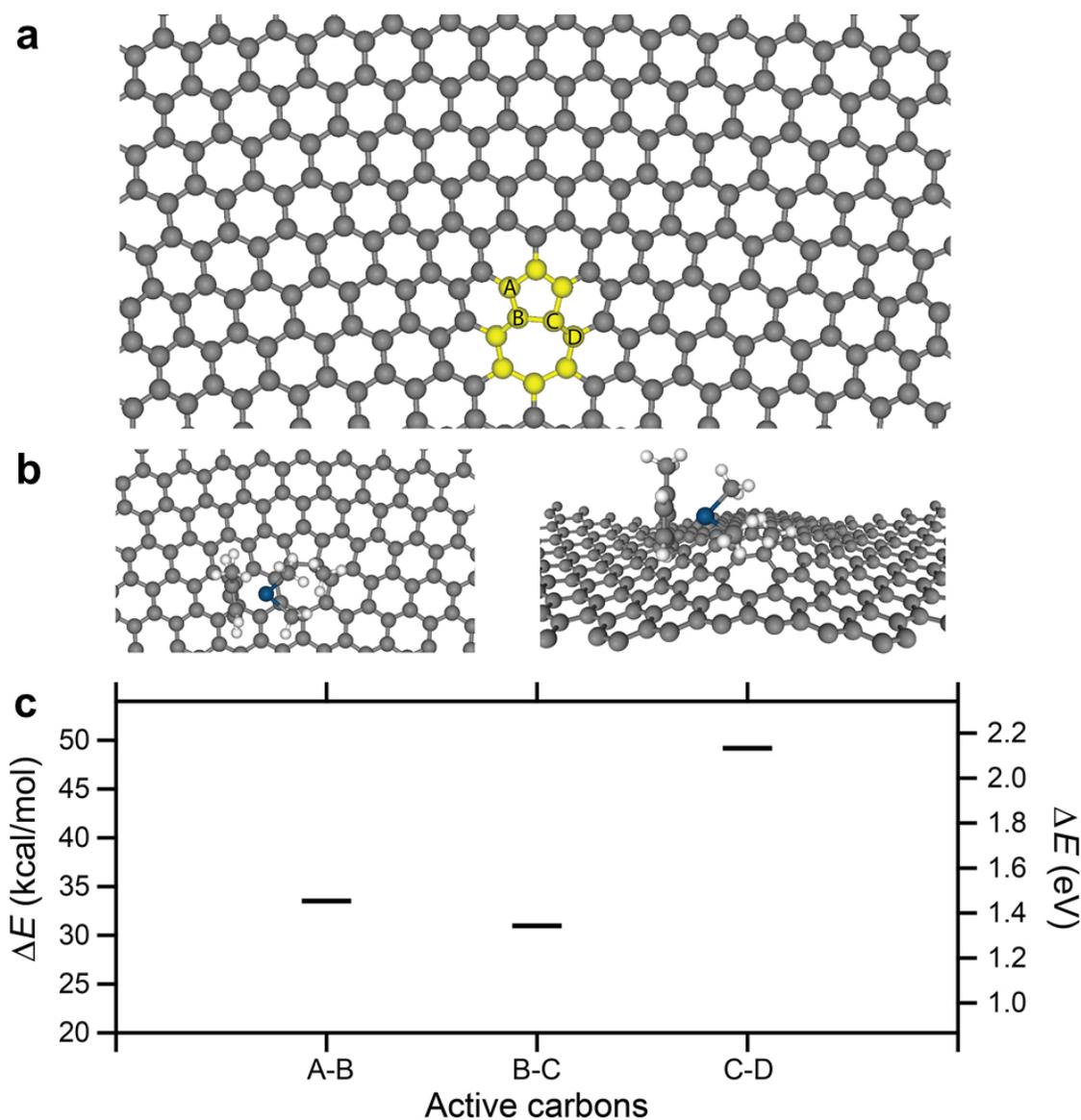

**Figure 4. DFT calculations of Pt-precursor chemisorption energetics on graphene grain boundary.** (**a**) Illustration of a periodic graphene grain boundary model with a heptagon-pentagon unit illustrated in yellow. Reactive carbon sites are labelled as A-D. (**b**) Top and side views of the surface species formed via chemisorption of $Pt(CH_3)_3CpCH_3$ on the 'B-C' site. (**c**) Reaction energies ($\Delta E$) for $C_2^* + Pt(CH_3)_3CpCH_3 \rightarrow C\text{-}Pt(CH_3)_2CpCH_3^* + C\text{-}CH_3^*$ as a function of the active $C_2^*$ site. The reaction energy is defined as the energy difference between the final chemisorbed product and the starting materials (grain boundary slab and gas phase $Pt(CH_3)_3CpCH_3$).

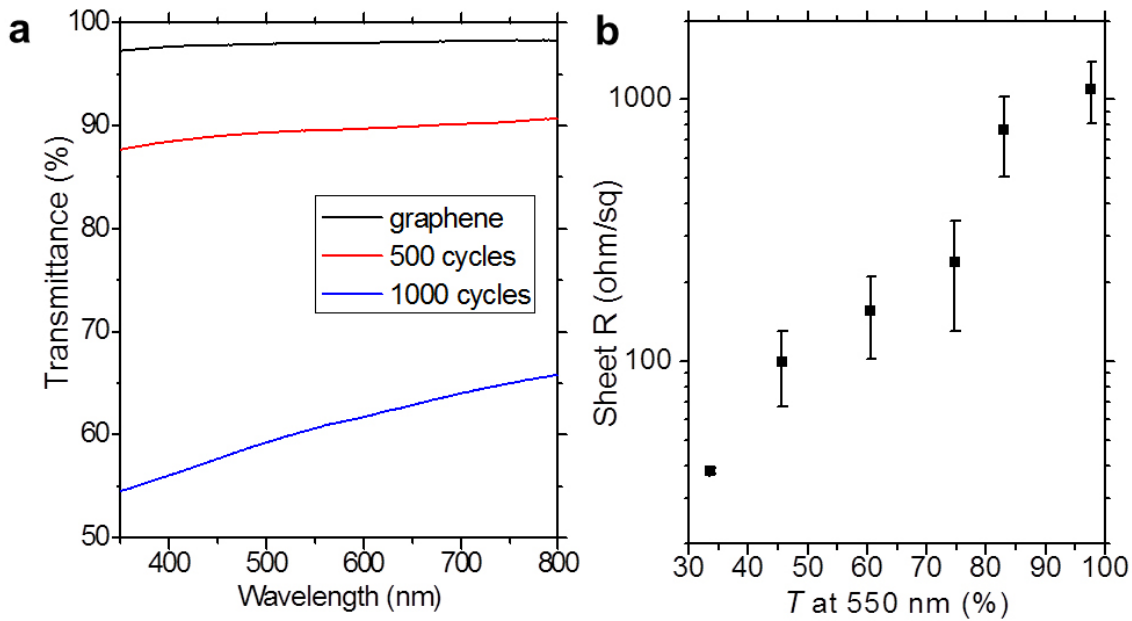

**Figure 5. Optical transmittance and sheet resistance measurement.** (**a**) Optical transmittance measurement of CVD graphene and Pt deposited samples. (**b**) Sample sheet resistance vs. optical transmittance at 550 nm wavelength for various Pt depositions. The error bars indicate the standard deviations from five samples.

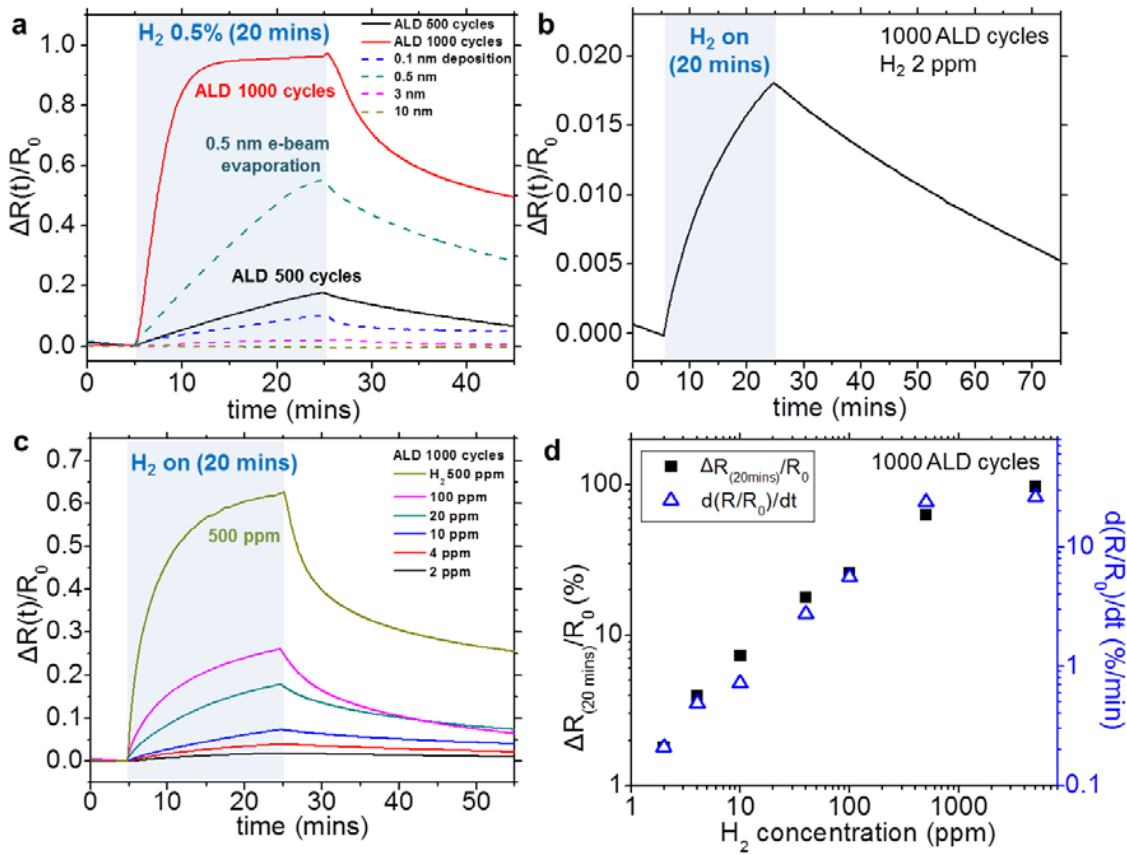

**Figure 6. $H_2$ sensing experiments with Pt ALD samples.** (**a**) The normalized resistance changes of Pt deposited graphene samples responding to 0.5 % concentration of $H_2$ gas. Pt-deposited graphene samples by ALD and e-beam evaporation with various thicknesses are compared. (**b**) The normalized resistance changes of an ALD sample (1000 cycles) responding to ultra-low $H_2$ concentration (2 ppm). (**c**) The normalized resistance changes of sample with 1000 ALD cycles responding to various $H_2$ concentrations. Note that the response to 2 ppm, which is shown in Figure 6b, is also plotted for comparison. (**d**) Log-scale plot of the normalized change of resistance (after 20 mins of $H_2$ exposure) and the initial rate of normalized resistance change as a function of $H_2$ concentration.

**Acknowledgements**

We thank Tae Hoon Lee and John To for valuable discussion and technical assistance. We also thank Prof. Oleg Yazyev for sharing a periodic graphene grain boundary model. This work was supported by Stanford Global Climate and Energy Program. Pt ALD was supported by the U.S. Department of Energy (DOE), Office of Science, Basic Energy Sciences (BES), under Award #DE-SC0004782. J.T.T. gratefully acknowledges the Academy of Finland (Grant 256800/2012) and the Finnish Cultural Foundation for financial support.

**Author contributions**

K.K. and H.B.R.L. contributed to this work equally. K.K., H.B.R.L., S.F.B., and Z.B. designed the experiments. H.B.R.L. and R.W.J. carried out Pt deposition by ALD on graphene samples. K.K. performed sample fabrication, characterizations, and gas sensing experiments. N.L., M.-G.K., C.P., and C.A. helped with sample fabrication and characterizations. J.T.T. performed theoretical calculations. K.K., H.B.R.L., J.T.T., and Z.B. co-wrote the paper. All authors discussed the results and commented on the manuscript.